\documentclass[11pt]{article}
\def\ben{\begin{equation}}
\def\een{\end{equation}}

\def\nn{\nonumber} \def\bd{\begin{document}} \def\ed{\end{document}}
\def\ds{\documentstyle} \let\fr=\frac \let\bl=\bigl \let\br=\bigr
\let\Br=\Bigr \let\Bl=\Bigl
\let\bm=\bibitem
\let\na=\nabla
\let\pa=\partial \let\ov=\overline
\newcommand{\be}{\begin{equation}}
\newcommand{\ee}{\end{equation}}
\def\ba{\begin{array}}
\def\ea{\end{array}}
\def\ft#1#2{{\textstyle{\frac{\scriptstyle #1}{\scriptstyle #2} } }}
\def\fft#1#2{{\frac{#1}{#2}}}
\def\del{\partial}
\def\vp{\varphi}
\def\sst#1{{\scriptscriptstyle #1}}
\def\oneone{\rlap 1\mkern4mu{\rm l}}
\def\td{\tilde}
\def\wtd{\widetilde}
\def\ie{{\it i.e.\ }}
\def\dalemb#1#2{{\vbox{\hrule height .#2pt
        \hbox{\vrule width.#2pt height#1pt \kern#1pt
                \vrule width.#2pt}
        \hrule height.#2pt}}}
\def\square{\mathord{\dalemb{6.8}{7}\hbox{\hskip1pt}}}
\newcommand{\ho}[1]{$\, ^{#1}$}
\newcommand{\hoch}[1]{$\, ^{#1}$}
\newcommand{\bea}{\setlength\arraycolsep{2pt} \begin{eqnarray}}
\newcommand{\eea}{\end{eqnarray}}
\newcommand{\ra}{\rightarrow}
\newcommand{\lra}{\longrightarrow}
\newcommand{\Lra}{\Leftrightarrow}
\newcommand{\bp}{\tilde \beta^\prime}
\newcommand{\tr}{{\rm tr} }
\newcommand{\Tr}{{\rm Tr} }
\def\0{{\sst{(0)}}}
\def\1{{\sst{(1)}}}
\def\2{{\sst{(2)}}}
\def\3{{\sst{(3)}}}
\def\4{{\sst{(4)}}}
\def\5{{\sst{(5)}}}
\def\6{{\sst{(6)}}}
\def\7{{\sst{(7)}}}
\def\8{{\sst{(8)}}}
\def\m{{\sst{(m)}}}
\def\n{{\sst{(n)}}}
\def\cA{{{\cal A}}}
\def\cB{{{\cal B}}}
\def\cF{{{\cal F}}}
\def\cG{{{\cal G}}}
\def\cH{{{\cal H}}}
\def\tV{\widetilde V}
\def\tW{\widetilde W}
\def\tH{\widetilde H}
\def\tE{\widetilde E}
\def\tF{\widetilde F}
\def\tA{\widetilde A}
\def\im{{{\rm i}}}
\def\tY{{{\wtd Y}}}
\def\ep{{\epsilon}}
\def\vep{{\varepsilon}}
\def\bD{{{\bar D}}}
\def\R{{{\mathbb R}}}
\def\C{{{\mathbb C}}}
\def\H{{{\mathbb H}}}
\def\CP{{{\mathbb C}{\mathbb P}}}
\def\RP{{{\mathbb R}{\mathbb P}}}
\def\Z{{{\mathbb Z}}}
\def\bA{{{\mathbb A}}}
\def\bB{{{\mathbb B}}}
\def\bC{{{\mathbb C}}}
\def\bD{{{\mathbb D}}}
\def\bE{{{\mathbb E}}}
\def\bZ{{{\mathbb Z}}}
\def\Re{{{\frak{Re}}}}
\def\Im{{{\frak{Im}}}}
\def\cosec{{\,\hbox{cosec}\,}}
\def\Gm{{\Gamma_{\!\! -}}}
\def\Gp{{\Gamma_{\!\! +}}}
\def\stan{{standard }}
\def\nonstan{{supernumerary }}
\def\p{{\partial}}
\def\kdel#1{{\fft{\del}{\del#1}}}

\def\bog{{Bogomolny }}
\def\om{{\omega}}

\newcommand{\nnr}{\nonumber \\}
\newcommand{\pd}{\partial}
\newcommand{\ud}{\textrm{d}}
\newcommand{\dTH}{T^{\prime \, 0}_\textrm{H}}
\newcommand{\dOi}{\Omega^{\prime \, 0}_i}
\newcommand{\bx}{{\bf x}}

\newcommand{\tamphys}{\it George and Cynthia Woods Mitchell  Institute
for Fundamental Physics and Astronomy,\\
Texas A\&M University, College Station, TX 77843-4242, USA}

\newcommand{\auth}{H. L\"u\hoch{\dagger\star}, 
Jianwei Mei\hoch{\dagger} and C.N. Pope\hoch{\dagger,\ddagger}
}

\begin{document}

\begin{flushright}
\hfill{
MIFP-09-18\ \ \ \ \ \ \ \  }\\
 %\hfill{
%\bf hep-th/yymmnnn}
\end{flushright}

\vspace{25pt}
\begin{center}
{\large {\bf Solutions to Ho\v{r}ava Gravity}}

\vspace{15pt}
\auth

\vspace{10pt}
\hoch{\dagger}{\tamphys}

\vspace{10pt}

%\hoch{\ddagger}{\it Interdisciplinary Center of Theoretical Studies,
%USTC, Hefei, Anhui 230026, PRC}

\hoch{\star}{\it Division of Applied Mathematics and Theoretical
Physics,\\
China Institute for Advanced Study,\\
Central University of Finance and Economics, Beijing, 100081, China
}

\vspace{10pt}

\hoch{\ddagger}{\it  DAMTP, Centre for Mathematical Sciences,
 Cambridge University,\\  Wilberforce Road, Cambridge CB3 OWA, UK}

\vspace{10pt}

%\hoch{\diamondsuit}{\it Physics Department,\\ 
%New York City College of Technology,\\ the City University of New York, 
%Brooklyn, NY 11201, USA}

\vspace{30pt}

\underline{ABSTRACT}
\end{center}

  Recently Ho\v{r}ava proposed a non-relativistic renormalisable theory
of gravitation, which reduces to Einstein's general relativity at
large distances, and that may provide a candidate for a UV completion
of Einstein's theory.  In this paper, we derive the full set of
equations of motion, and then we obtain spherically symmetric
solutions and discuss their properties.  We also obtain solutions for
the Friedman-Lema\^itre-Robertson-Walker cosmological metric.

\vspace{15pt}

\thispagestyle{empty}

%\pagebreak
%\voffset=0pt
%\setcounter{page}{1}

%\tableofcontents

%\addtocontents{toc}{\protect\setcounter{tocdepth}{2}}

%%%%%%%%%%%%%%%%%%%%%%%%%%%%%%%%%%%%%%%%

\newpage

Recently a new four-dimensional theory of gravity was proposed by
Ho\v{r}ava \cite{horava09}, inspired by condensed matter
models of dynamical critical systems. It has manifest three-dimensional
spatial general covariance and time-reparameterisation invariance, but
only acquires four-dimensional general covariance in an infra-red large
distance limit.  It may be described in a language akin to the ADM 3+1 
dimensional formulation of general relativity, but in which Einstein gravity
is modified so that the full underlying four-dimensional covariance is
broken.  The
nature of the modifications is governed by a rather strong principle 
of ``detailed balance'' \cite{horava09}.

   In the ADM
formalism, the four-dimensional metric of general relativity 
is parameterised as
\cite{adm}
%%%
\be
ds_4^2= - N^2  dt^2 + g_{ij} (dx^i - N^i dt) (dx^j - N^j dt)\,,
\label{metricans}
\ee
%%%%
where the lapse, shift and 3-metric $N$, $N^i$ and $g_{ij}$ are all
functions of $t$ and $x^i$.  In the simplest version of the theory
in \cite{horava09}, the lapse function $N$ is viewed as a gauge field for
time reparameterisations, and it is effectively restricted to depend
only on $t$, but not the spatial coordinates $x^i$.  A closer parallel 
with general relativity is achieved if this ``projectability'' restriction is
relaxed. Thus one may take a broader view of the Ho\v{r}ava proposal as
a class of theories in which the relative coefficients of the terms in the
ADM decomposition of the Einstein-Hilbert action are modified, and additional
terms involving higher spatial derivatives are included too.  The higher
derivative terms can improve the renormalisability of the theory, without
the usual attendant problems of ghosts that would arise if higher time
derivatives were present too.  In this letter, we shall take the broader
viewpoint (discussed also in \cite{horava09}) in which the theory is 
essentially a class of 3+1 dimensional modifications of general relativity. 
 
   The ADM decomposition (\ref{metricans}) of the
Einstein-Hilbert action is given by
%%%
\be
S_{EH} = \fft{1}{16\pi G} \int d^4x \sqrt{g} N (K_{ij} K^{ij} - K^2 + R -
2\Lambda)\,,\label{ehlag}
\ee
%%%%
where $G$ is Newton's constant and $K_{ij}$ is defined by
%%%
\be
K_{ij} = \fft{1}{2N} (\dot g_{ij} - \nabla_i N_j - \nabla_j N_i)\,.
\ee
%%%%
Here, a dot denotes a derivative with respect to $t$.

   The action of the 
theory proposed by Ho\v{r}ava \cite{horava09} can be written as
%%%
\bea%
S&=&\int dtd^3\bx\, ({\cal L}_0 + {\cal L}_1)\,,\nn\\ {\cal L}_0
&=& \sqrt{g}N\left\{\frac{2}{\kappa^2}(K_{ij}K^{ij} -\lambda
K^2)+\frac{\kappa^2\mu^2(\Lambda_W R
  -3\Lambda_W^2)}{8(1-3\lambda)}\right\}\,,\nn\\ {\cal L}_1&=&
\sqrt{g}N\left\{\frac{\kappa^2\mu^2 (1-4\lambda)}{32(1-3\lambda)}R^2
-\frac{\kappa^2}{2w^4} \left(C_{ij} -\frac{\mu w^2}{2}R_{ij}\right)
\left(C^{ij} -\frac{\mu w^2}{2}R^{ij}\right)\right\}\,,\label{action}
\eea% 
%%%%%
where $\lambda\,,\kappa\,,\mu\,,w$ and $\Lambda_W$ are constant
parameters, and $C_{ij}$ is the Cotton tensor, defined by
%%%
\be
C^{ij}=\epsilon^{ik\ell}\nabla_k\left(R^j{}_\ell
-\frac14R\delta_\ell^j\right) \,.\label{def.K.C}
\ee
%%%%
Comparing ${\cal L}_0$ to that of general relativity in the ADM
formalism, the speed of light, Newton's constant and the cosmological
constant emerge as
%%%%
\be
c=\fft{\kappa^2\mu}{4} \sqrt{\fft{\Lambda_W}{1-3\lambda}}\,,\qquad
G=\fft{\kappa^2}{32\pi\,c}\,,\qquad
\Lambda=\ft32 \Lambda_W\,.\label{cg}
\ee
%%%
(One can without loss of generality choose units so that $c=1$. Indeed
this was done in (\ref{metricans}) and (\ref{ehlag}).)
Furthermore, the requirement that ${\cal L}_0$ be equivalent to the
usual Einstein-Hilbert Lagrangian, and thus have
four-dimensional general covariance, implies that one must take
$\lambda=1$.  In Ho\v{r}ava gravity, $\lambda$ represents a dynamical
coupling constant, susceptible to quantum corrections \cite{horava09}.
In order to recover general relativity in the infra-red limit, the
other dynamical coupling constants would need to flow so that
$\mu\rightarrow 0$, $\Lambda_W\rightarrow \infty$ and $w\rightarrow\infty$, 
with $\mu^2 \Lambda_W$ fixed.  (There would also need to be a fine-tuning
dynamical
mechanism, presumably from a matter sector, 
to subtract the now-infinite cosmological constant 
proportional to $\mu^2 \Lambda_W^2$.)
 Note from (\ref{cg}) that for $\lambda> \ft13$, the cosmological
constant $\Lambda_W$ has to be negative.  The cosmological
implications of the action (\ref{action}) were recently discussed in
\cite{c1,c2}.  See also \cite{c3}.

  The constants $\mu$ and $w^2$ are real, and have their origin as the
Newton constant and Chern-Simons coupling of Euclideanised
three-dimensional topologically massive gravity \cite{djt}.  However,
if we make an analytic continuation of these parameters, namely
$\mu\rightarrow {\rm i} \mu$, $w^2 \rightarrow -{\rm i} w^2$,
the four-dimensional action remains real, with the sign of all terms 
except $(K_{ij}K^{ij}-\lambda K^2)$ in (\ref{action}) now being reversed.
In this case, the emergent speed of light becomes
$c=\ft14 \kappa^2\, \mu\sqrt{\Lambda_W/(3\lambda-1)}$.
The requirement that this speed be real implies that $\Lambda_W$ be
positive for $\lambda>\ft13$.

We now consider the equations of motion for the action (\ref{action}).
The equation following from
the variation of $N$, is purely algebraic, and is given by
%%%%
\be
\fft{2}{\kappa^2}(K_{ij}K^{ij} -\lambda K^2) -
\frac{\kappa^2\mu^2(\Lambda_W R
-3\Lambda_W^2)}{8(1-3\lambda)} -\frac{\kappa^2\mu^2
(1-4\lambda)}{32(1-3\lambda)}R^2 + \frac{\kappa^2}{2w^4}
Z_{ij} Z^{ij}=0\,,\label{eom1}
\ee%
%%%
where
%%%
\be
Z_{ij}\equiv C_{ij} - \fft{\mu w^2}{2} R_{ij}\,.
\ee
%%%%%
The variation $\delta N^i$ implies
%%%
\be
\nabla_k(K^{k\ell}-\lambda\,Kg^{k\ell})=0\,.\label{eom2}
\ee%
%%%
The equations of motion due to the variation of
$\delta g^{ij}$ are more complicated; they are given by
%%%%
\be%
\frac{2}{\kappa^2}E_{ij}^\1-\frac{2\lambda}{\kappa^2}E_{ij}^\2
+\frac{\kappa^2\mu^2\Lambda_W}{8(1-3\lambda)}E_{ij}^\3
+\frac{\kappa^2\mu^2(1-4\lambda)}{32(1-3\lambda)}E_{ij}^\4
-\frac{\mu\kappa^2}{4w^2}E_{ij}^\5
-\frac{\kappa^2}{2w^4}E_{ij}^\6=0\,,\label{eom3}\ee%
%%%
where
\bea%
E_{ij}^\1&=&
N_i \nabla_k K^k{}_j + N_j\nabla_k K^k{}_i -K^k{}_i \nabla_j N_k-
   K^k{}_j\nabla_i N_k - N^k\nabla_k K_{ij}\nn\\
&& - 2N K_{ik} K_j{}^k
  -\frac12 N K^{k\ell} K_{k\ell}\, g_{ij} + N K K_{ij} + \dot K_{ij}
\,,\nn \\
E_{ij}^\2&=& \frac12 NK^2 g_{ij}+ N_i \pd_j K+
N_j \pd_i K- N^k (\pd_k K)g_{ij}+  \dot K\, g_{ij}\,\,,\nn\\
E_{ij}^\3&=&N(R_{ij}-
\frac12Rg_{ij}+\frac32\Lambda_Wg_{ij})-(
\nabla_i\nabla_j-g_{ij}\nabla_k\nabla^k)N\,,\nn\\
E_{ij}^\4&=&NR(2R_{ij}-\frac12Rg_{ij})-
2 \big(\nabla_i\nabla_j
-g_{ij}\nabla_k\nabla^k\big)(NR)\,\,,\nn\\
E_{ij}^\5&=&\nabla_k\big[\nabla_j(N Z^k_{~~i})
+\nabla_i(N Z^k_{~~j})\big]  -\nabla_k\nabla^k(NZ_{ij})
-\nabla_k\nabla_\ell(NZ^{k\ell})g_{ij}\,\,,\nn\\
E_{ij}^\6&=&-\frac12NZ_{k\ell}Z^{k\ell}g_{ij}+
2NZ_{ik}Z_j^{~k}-N(Z_{ik}C_j^{~k}+Z_{jk}C_i^{~k})
+NZ_{k\ell}C^{k\ell}g_{ij}\nn\\
&&-\frac12\nabla_k\big[N\epsilon^{mk\ell}
(Z_{mi}R_{j\ell}+Z_{mj}R_{i\ell})\big]
+\frac12R^n{}_\ell\, \nabla_n\big[N\epsilon^{mk\ell}(Z_{mi}g_{kj}
+Z_{mj}g_{ki})\big]\nn\\
&&-\frac12\nabla_n\big[NZ_m^{~n}\epsilon^{mk\ell}
(g_{ki}R_{j\ell}+g_{kj}R_{i\ell})\big]
-\frac12\nabla_n\nabla^n\nabla_k\big[N\epsilon^{mk\ell}
(Z_{mi}g_{j\ell}+Z_{mj}g_{i\ell})\big]\nn\\
&&+\frac12\nabla_n\big[\nabla_i\nabla_k(NZ_m^{~n}\epsilon^{mk\ell})
g_{j\ell}+\nabla_j\nabla_k(NZ_m^{~n}\epsilon^{mk\ell})
g_{i\ell}\big]\nn\\
&&+\frac12\nabla_\ell\big[\nabla_i\nabla_k(NZ_{mj}
\epsilon^{mk\ell})+\nabla_j\nabla_k(NZ_{mi}
\epsilon^{mk\ell})\big]-\nabla_n\nabla_\ell\nabla_k
(NZ_m^{~n}\epsilon^{mk\ell})g_{ij}\,.\eea%
%%%%
(The equations of motion were also obtained in \cite{c2}.) Note that
in deriving these equations of motion, we have relaxed the ``projectability''
restriction and allowed the lapse function $N$ to depend on the
spatial coordinates $x^i$ as well as $t$.  Had we not done so, equation
(\ref{eom1}) would instead have to hold only when integrated over all
space.  Obtaining general relativity in an infra-red limit
could then be problematical.

    We may now seek
static, spherically symmetric solutions with the metric ansatz
%%%%
\be
ds^2 = - N(r)^2\,dt^2 + \fft{dr^2}{f(r)} + r^2 (d\theta^2
+\sin^2\theta d\phi^2)\,.
\ee
%%%%
   If we consider a system with the Lagrangian ${\cal L}_0$ only,
we obtain the (A)dS Schwarzschild black hole with
%%%
\be
N^2=f=1 - \fft{1}{2} \Lambda_W\, r^2 - \fft{M}{r}\,.\label{adsbh}
\ee
%%%
The easiest way to obtain the solution for the full Lagrangian ${\cal
L}_0 + {\cal L}_1$ is to substitute the metric ansatz into the action,
and then vary the functions $N$ and $f$.  This is a valid procedure
since the ansatz contains all the allowed singlets compatible with the
$SO(3)$ action on the $S^2$.  The resulting reduced Lagrangian, up to
an overall scaling constant, is given by
%%%%
\be
{\cal L}= \fft{N}{\sqrt{f}}\Big(2 - 3\Lambda_W r^2 -2 f - 2r f' +
\fft{\lambda-1}{2\Lambda_W} f'^2 - \fft{2\lambda (f-1)}{\Lambda_W r}f'
+ \fft{(2\lambda-1)(f-1)^2}{\Lambda_W r^2}\Big).
\ee
%%%%%
There are in total three solutions.\footnote{We have also verified
that all the solutions indeed satisfy the full set of equations of
motion (\ref{eom1}), (\ref{eom2}) and (\ref{eom3}).}  The first
solution is given by
%%%
\be
f=1+x^2\,,\qquad x=\sqrt{-\Lambda_W}\, r\,.\label{fsol1}
\ee
%%%%
This is valid for all $\lambda$, but strangely enough, the function
$N(r)$ is unconstrained.  This suggests that if we have a hyperbolic
spatial section, the Newtonian potential associated with $g_{tt}=-N^2$
can be an arbitrary function of $r$.  (In fact, with $f$ given by
(\ref{fsol1}), we have verified that $N$ can be an arbitrary function
of all the space-time coordinates.)  As we shall show later, this
particular feature is specific to the choice of coefficients that was
made in \cite{horava09} in order to satisfy the condition of
``detailed balance.''

   There are two more solutions, in which both $f$ and $N$ are
determined, given by
%%%%
\be%
f=1 + x^2-\alpha\, x^{\frac{2\lambda\pm\sqrt{6\lambda-2}}{\lambda-1}}
\,,\quad
N=x^{-\frac{1+3\lambda\pm2\sqrt{6\lambda-2}}{\lambda-1}}
\sqrt{f}\,,\label{solution1}
\ee
%%%%
where $\alpha$ is an integration constant.  For the solution to be
real, it is necessary that $\lambda >1/3$.  It follows from (\ref{cg})
that $\Lambda_W$ is a negative cosmological constant if we consider
the action (\ref{action}); $\Lambda_W$ is a positive cosmological
constant if we consider instead the action with the continuation
$\mu\rightarrow {\rm i} \mu$, $w^2\rightarrow -{\rm i} w^2$.
In the limit where $\lambda=1/3$, the function $f$ becomes that of the
(A)dS black hole (\ref{adsbh}), but with twice the cosmological
constant.  
The solution has a curvature singularity at $x=0$ for general
$\lambda$.  It also has a curvature singularity at $x=\infty$ if
$(2\pm\sqrt{6\lambda-2})(\lambda-1)>0$.

     It is of particular interest to investigate the $\lambda=1$
solution, in which case, the functions $f$ and $N$ are given by
%%%%
\be
N^2=f=1 + x^2 - \alpha\, \sqrt{x}\,.\label{nf}
\ee
%%%
The solution is asymptotically AdS$_4$, with a horizon at $x=x_+$,
where $x_+$ is the largest root of $f$.  The temperature is given by
%%%
\be
T=\fft{(3x_+^2-1)\sqrt{-\Lambda_W}}{8\pi x_+}\,.
\ee
%%%
There exists an extremal limit in which $\alpha=4/3^{3/4}$, with the
horizon located at $x=1/\sqrt3$, for which the temperature
vanishes.  The solution interpolates between the AdS$_2\times S^2$ at
the horizon and AdS$_4$ at asymptotic infinity.  The significant
difference between the solution (\ref{nf}) and the usual AdS
Schwarzschild black hole (\ref{adsbh}) suggests that general
relativity is not always recovered at large distance.\footnote{
It was observed in \cite{svw} that by writing
the Schwarzschild metric in Painlev\'e-Gullstrand coordinates, it
does in fact satisfy
the projectability condition (with $N(t)=1$).
We find that the AdS-Schwarzchild black hole can also be written
in Painlev\'e-Gullstrand type coordinates, and it is given by
%%%%
\be
ds^2=-dt^2 + \Big(dr + \sqrt{\fft{M}{r} +
\fft{\Lambda_W}2  r^2}\,  dt\Big)^2 + r^2 d\Omega^2\,.
\ee
%%%
Since in this coordinate system $g_{ij}$ is flat, it follows
that the metric, with no modifications,  gives an exact solution to
the Ho\v rava theory with $\lambda=1$, since the
higher-order derivative corrections, from ${\cal L}_1$, vanish.
However, owing to the negative
cosmological constant $\Lambda_W$, the solution has no asymptotic
$r\rightarrow \infty$ region. (Recall that one no longer has the
freedom of four-dimensional general covariance.)}

The action (\ref{action}) was obtained by imposing the condition of
detailed balance \cite{horava09}.  We may, however, entertain the idea
of deviating slightly from detailed balance, by considering the
Lagrangian
%%%
\be
{\cal L}= {\cal L}_0 + (1 - \epsilon^2) {\cal L}_1\,.
\ee
%%%
There exist two pure AdS$_4$ solutions.  The function $f$ is given by
%%%
\be
f=1 - \fft{\Lambda_W}{1 + \epsilon} r^2\,,
\ee
%%%
where $\epsilon$ can take both positive and negative values.
The remaining equations imply that
%%%
\be
\epsilon (\Lambda_W r\, N + (1 + \epsilon - \Lambda_W r^2) N')=0\,.
\ee
%%%
Thus we see that when detailed balance is satisfied, corresponding to
$\epsilon=0$, the function $N$ is unconstrained, as previously noted.
For $\epsilon\ne0$, we find that $N^2=f$, giving rise to the AdS$_4$
spacetimes.

 The general solution for $f$ and $N$ can also be obtained, in the
case of non-vanishing $\epsilon$.  Owing to its complexity, we shall
present only the special case where $\lambda=1$, for which it is given
by
%%%
\be
N^2=f= 1 + \fft{x^2}{1-\epsilon^2} -
\fft{\sqrt{\alpha^2(1-\epsilon^2) x + \epsilon^2 x^4}}{1 -\epsilon^2}
\,.
\ee
%%%
The large distance behaviour of the function $f$ is given by
%%%
\be
f=1 + \fft{x^2}{1+\epsilon} - \fft{\alpha^2}{2\epsilon\, x}
+ {\cal O}(x^{-4})\,.
\ee
%%%
Thus we see that for non-vanishing $\epsilon$, the metric has finite
mass, which becomes divergent for the detailed-balance value
$\epsilon=0$, in which case the function $f$ becomes the one given in
(\ref{nf}).  When $\epsilon=1$, the solution becomes the
AdS-Schwarzschild black hole (\ref{adsbh}).

We may also look for cosmological solutions of the
Friedman-Lema\^itre-Robertson-Walker form
%%%%
\be%
ds^2= - dt^2+a^2(t)\bigg[\frac{dr^2}{1-kr^2}+
r^2(d\theta^2+\sin^2\theta d\phi^2)\bigg]\,,\ee%
%%%
where $k=1,0,-1$ corresponding to a closed, flat or open universe
respectively. Supposing that the matter contribution is equivalent to
an ideal fluid, we find that
%%%%
\bea%
\left(\frac{\dot a}{a}\right)^2&=&\frac{2}{3\lambda-1}
\left(\frac{\Lambda_W}{2}
+\frac{8\pi G_N\rho}{3}-\frac{k}{a^2}+\frac{k^2}{2\Lambda_W a^4}
\right)\,,
\label{fried.1}\\
\frac{\ddot a}{a}&=&\frac{2}{3\lambda-1}\left(\frac{\Lambda_W}{2}-
\frac{4\pi G_N}{3}(\rho+3p)-\frac{k^2}{2\Lambda_W a^4}\right)\,.
\label{fried.2}\eea%
%%%%
It is interesting to note that for $k=0$, there is no contribution
from the higher-order derivative terms in the action.  For $k\ne 0$,
these contributions becomes dominant for small $a$, but weak at large
$a$, implying that the cosmological solutions of general relativity
are recovered at large scales.

      For vacuum solutions with $p=\rho=0$, we have
%%%
\be
\left(\frac{\dot a}{a}\right)^2=\frac{\Lambda_W}{3\lambda-1}
\Big(1 - \fft{k}{\Lambda_W a^2}\Big)^2\,.
\ee
%%%
It follows from (\ref{cg}) that for the action (\ref{action}), the
right-hand side of the equation is negative definite for
$\lambda>\ft13$.  Thus in this case, solutions only exist when $k=-1$,
in which case $a$ is a constant given by $a^2=-1/\Lambda_W$.

We can instead consider the action which is obtained
from the analytic continuation $\mu\rightarrow {\rm i} \mu$, 
$w^2\rightarrow -{\rm i} w^2$ we discussed earlier.
The reality condition for the speed of light now implies
that $\Lambda_W$ is positive for $\lambda>\ft13$.  The solution is
given by
%%%%
\be
a^2 = \fft{k}{\Lambda_W} + \alpha\, 
e^{\ft{2\sqrt{\Lambda_W}}{\sqrt{3\lambda-1}}\,\,t}\,.
\ee
%%%
It is of interest to note that for $k=1$, there is a minimum
scaling factor $a_{\rm min}=1/\sqrt{\Lambda_W}$.

In this paper, we have studied the recently-proposed non-relativistic
and renormalisable gravity theory introduced in \cite{horava09}.  We
derived the full set of equations of motion, and then considered the
static, spherically-symmetric solutions. We found that there exists a
solution where the spatial section is a hyperbolic space and the
metric component $g_{tt}=-N^2$ can be an arbitrary function of all the
space-time coordinates.  We demonstrated that this feature occurs
because certain coefficients in \cite{horava09} are chosen to satisfy
a condition of detailed balance.  The system also admits AdS$_2\times
S^2$ vacuum solutions.  In addition, there exist black hole solutions
that interpolate between AdS$_2\times S^2$ at the horizon and AdS$_4$
at asymptotic infinity.  The asymptotic fall-off of the metric for the
black hole solutions is much slower than that of the AdS-Schwarzschild
black hole in general relativity, suggesting that Einstein's gravity
does not always appear to be recovered at large distance.  We also
obtained cosmological vacuum solutions.

\section*{Acknowledgement}

We are grateful to Dick Arnowitt and 
Juan Maldacena for useful discussions, and to
George Mitchell and Sheridan Lorenz for hospitality at Cook's Branch
Conservancy where this work was completed.  C.N.P. is supported in
part by DOE grant DE-FG03-95ER40917.

\end{document}